\documentclass{PoS}

\usepackage{amsmath,amssymb}

\title{Temperature dependence of electrical conductivity and dilepton rates from hot
quenched lattice QCD}

\ShortTitle{T-dependence of electrical conductivity and dilepton rates}

\author{\speaker{Olaf Kaczmarek} and Marcel M\"uller
\\
   Fakult\"{a}t f\"{u}r Physik,
 Universit\"{a}t Bielefeld, D-33615 Bielefeld, Germany\\
        E-mail: \email{okacz@physik.uni-bielefeld.de}}

\abstract{We present new results on the continuum extrapolation of the vector current
correlation function in the deconfined phase for three temperatures close to
the critical temperature utilizing quenched clover improved Wilson fermions and
light quark masses.
A systematic analysis on multiple lattice spacing allows to perform the
continuum limit of the correlation function and to extract spectral properties
in the continuum limit. These results provide constraints for the electrical
conductivity and the thermal dilepton rates in the quark gluon plasma for the
given temperature range. In addition results on the continuum
extrapolation at finite momenta related to thermal photon rates are presented.}

\FullConference{31st International Symposium on Lattice Field Theory - LATTICE 2013\\
		July 29 - August 3, 2013\\
		Mainz, Germany}

\begin{document}

\section{Introduction}
Thermally produced dileptons and photons are important experimental observables
to study the the quark gluon plasma medium produced in current
heavy-ion-experiments at LHC and RHIC \cite{Rapp:2009yu}. Both quantities are related to the
vector meson spectral function and therefore indirectly to the vector correlation
function. The energy regime of the current experimental studies
requires non-perturbative ab initio lattice QCD calculations. The results
presented in this article extend our previous studies
\cite{Ding:2010ga,Francis:2011bt,Ding:2013qw} where we
performed the continuum extrapolation of the vector meson correlation function
for the first time at a temperature of $T/T_c=1.45$ in the quenched
approximation using non-perturbatively clover improved Wilson fermions.
Here we will discuss the temperature dependence of the electrical conductivity
and dilepton rates obtained from continuum extrapolated correlation functions
at temperatures of 1.1, 1.2 and 1.4~$T_c$.

%\begin{figure}
%\begin{center}
%\includegraphics{V1234_FinalRatioFreeLat_Tc110.pdf}
%\includegraphics{V1234_FinalRatioFreeLat_Tc120.pdf}
%\caption{Vector meson correlation function $G_V(\tau T)$ at $1.1~T_c$~(left)
%  and $1.2~T_c$~(right) for the three lattice spacings and continuum extrapolation.}
%\label{plot:V1234}
%\end{center}
%\end{figure}
\begin{figure}
\begin{center}
\includegraphics{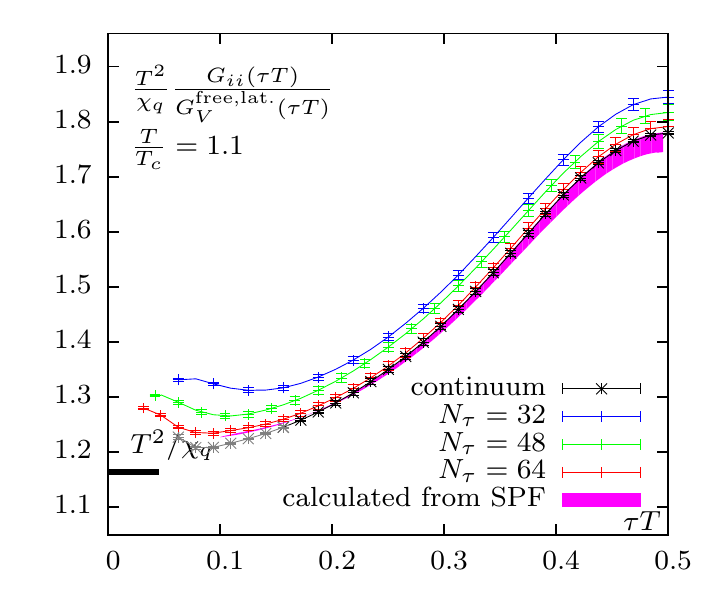}
\includegraphics{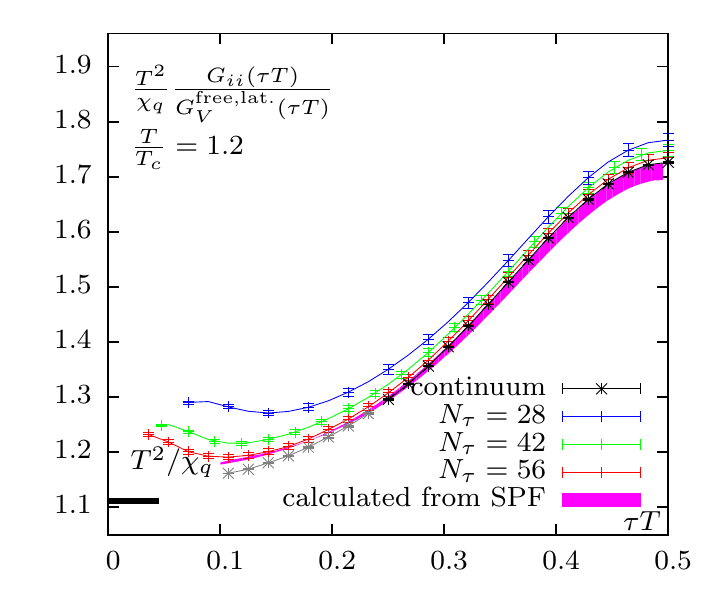}
\caption{Vector meson correlation function $G_{ii}(\tau T)$ at $1.1~T_c$~(left)
  and $1.2~T_c$~(right) for the three lattice spacings and the continuum
  extrapolation. The band shows the result and systematic error of the fit
  using our Ansatz for the spectral function.}
\label{plot:V1234c}
\end{center}
\end{figure}

\section{Vector correlation function}

The Euclidean time two-point correlation function $G(\tau,\vec p)$ 
of the vector current $J_\mu$
is a quantity directly accessible in lattice QCD calculations,

\begin{equation}
G_{\mu\nu}(\tau,\vec p) = \int d^3x  J_\mu(\tau,\vec x) J_\nu^\dag(0,\vec0)  e^{i\vec p \vec x} 
 \quad \text{with} \quad J_\mu(\tau,\vec x) = \bar q(\tau, \vec x) \gamma_\mu q(\tau,\vec x).
\end{equation}

Only contributions of quark line connected diagrams are included. Disconnected diagrams
cause a high numerical effort and are expected to be small in the high temperature phase of QCD
\cite{Allton:2005gk, Gavai:2001ie}.
The correlation function
directly relates to the spectral function via 

\begin{equation}
G_H(\tau, \vec p,T) = \int_0^\infty \frac{d\omega}{2\pi} \rho_H(\omega, \vec p, T) 
\frac{\cosh(\omega(\tau - 1/2 T))}{\sinh{(\omega/2T)}} \quad \text{with:} \quad H=00, ii, V.
\label{spfToCorr}
\end{equation}

\begin{figure}[thbp]
\begin{center}
\includegraphics{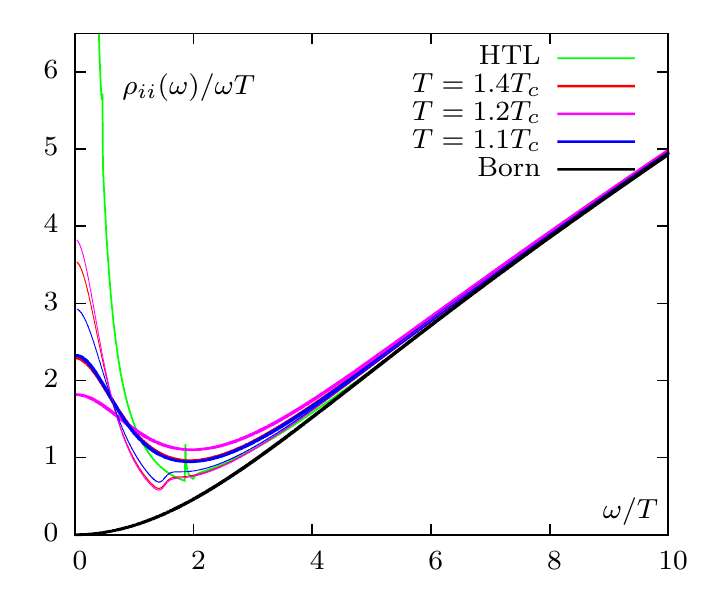}
\includegraphics{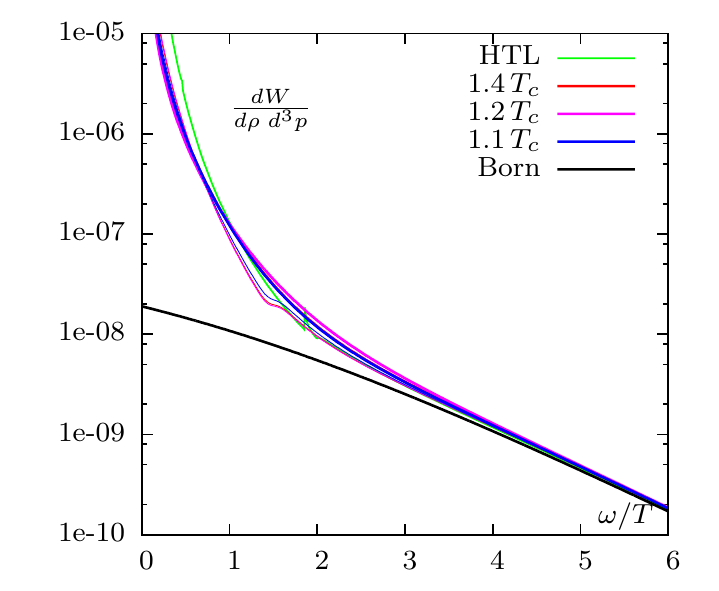}
\caption{
Results for the spectral functions~(left) and thermal dilepton rates
(right) calculated from the fit results to our spectral function Ansatz for three
temperatures.}
\label{plot:spf}
\end{center}
\end{figure}

Here $\rho_{ii}$ denotes a sum over the spatial components and $\rho_{00}$ denotes
the time-like components which is related to the quark number
susceptibility. The full vector spectral function 
is denoted by $\rho_V = \rho_{00} + \rho_{ii}$. 

\subsection{Ansatz for the spectral function}

The time-like component of the vector correlator $G_{00}$ and thereby the corresponding
spectral function $\rho_{00}$ is related to 
the quark number susceptibility $\chi_q$. Since the quark 
number is conserved, the correlator is constant in 
(here Euclidean) time, $G_{00}(\tau T) = - \chi_q T$
and its spectral representation is given by a delta function

\begin{equation}
\rho_{00}(\omega) = -2 \pi \chi_q \omega \delta(\omega).
\label{chiEquation}
\end{equation}

The spatial components of the spectral functions increase quadratically for large
values of $\omega$, in the free field limit for massless quarks to

\begin{equation}
\rho_{ii}^\text{free}(\omega) = 2 \pi T^2 \omega \delta(\omega) + \frac{3}{2\pi} \omega^2 \tanh(\omega/4T).
\end{equation}

In this limit, the delta peak in the spatial and the time-like component
of the spectral function cancel. However with interactions this is
not the case. The time-like component maintains a delta peak since
it is linked to the conserved current, but in the spatial component
the delta peak is smeared out and expected to be described by
a Breit-Wigner peak \cite{Aarts:2002cc,Moore:2006qn,Petreczky:2005nh, Hong:2010at},

\begin{equation}
\rho_{ii}^\text{interac.}(\omega) = \chi_q c_{\text{BW}} \frac{\omega \Gamma}{\omega^2 + (\Gamma/2)^2}
 + (1+\kappa) \frac{3}{2\pi} \omega^2 \tanh(\omega/4T).
\label{spfAnsatz}
\end{equation}

This phenomenologically inspired Ansatz leaves three parameters, the strength ($c_{\text{BW}}$) and width ($\Gamma$) of
the Breit-Wigner peak as well as $\kappa$, which accounts for the deviation from free theory
(see \cite{Burnier:2012ts,Rapp:2013nxa} for similar Ans\"atze).
The relation of this Ansatz to the correlator obtained on the lattice  
is given by \eqref{spfToCorr}.

\begin{figure}
\begin{center}
\includegraphics[width=8cm]{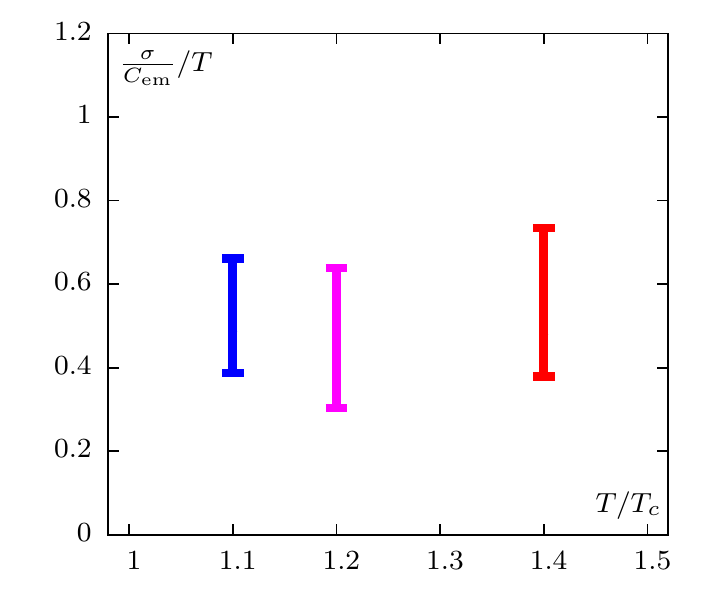}
\caption{Temperature dependence of the electrical conductivity. Errors are
  obtained from the systematic error analysis (for details see [2]).}
\label{plot:cond}
\end{center}
\end{figure}

The fits are not performed directly to the correlation function $G_{ii}$ 
but the ratio of the correlation function 
normalized by the quark number susceptibility (as given in 
\eqref{chiEquation}) to become independent of any
renormalization and by the free field
correlation function $G_V^\text{free}(\tau T)$, yielding
a smooth function that does not fall off over multiple decades
like the correlation function. 
Furthermore, due to asymptotic freedom, the correlation function
should approach the non-interacting limit at asymptotically small distances.
The spectral function is thereby fitted to reproduce

\begin{equation}
\frac{ G_{ii}(\tau T) /  G_{00} }{ 
G_V^\text{free}(\tau T) / G_{00}^\text{free}  }.  
\label{fitRatios}
\end{equation}

\begin{table}[h]
\begin{center}
\begin{tabular}{|c|c|c|c|c|c|c|c|}
\hline
$T/T_c$ & $N_\tau$ & $N_\sigma$ & $\beta$ & $\kappa$ & $ 1/a $ [GeV] & $  a $[fm] & \#conf \\
\hline
    & 32 & 96 & 7.192  & 0.13440 & 10.4 & 0.019 & 314 \\ 
1.1 & 48 & 144 & 7.544 & 0.13383 & 15.5 & 0.013 & 367 \\
    & 64 & 192 & 7.793 & 0.13345 & 20.4 & 0.010 & 242 \\ 
\hline
    & 28 & 96 & 7.192  & 0.13440 & 10.4 & 0.019 & 232 \\ 
1.2 & 42 & 144 & 7.544 & 0.13382 & 15.5 & 0.013 & 417 \\
    & 56 & 192 & 7.793 & 0.13345 & 20.4 & 0.010 & 195 \\ 
\hline
    & 24 & 128 & 7.192 & 0.13440 & 10.4 & 0.019 & 232 \\ 
1.4 & 32 & 128 & 7.458 & 0.13383 & 14.1 & 0.014 & 417 \\
    & 48 & 128 & 7.793 & 0.13340 & 20.4 & 0.010 & 195 \\ 
\hline
\end{tabular}
\end{center}
\caption{Summary of simulation parameters}
\label{simulationParameters}
\end{table}

More details on the fit procedure and the error analysis can be found
in \cite{Ding:2010ga}.
Having obtained the spectral function, relevant properties
of the medium can be calculated, e.g. the electrical conductivity as

\begin{figure}[thbp]
\begin{center}
\includegraphics{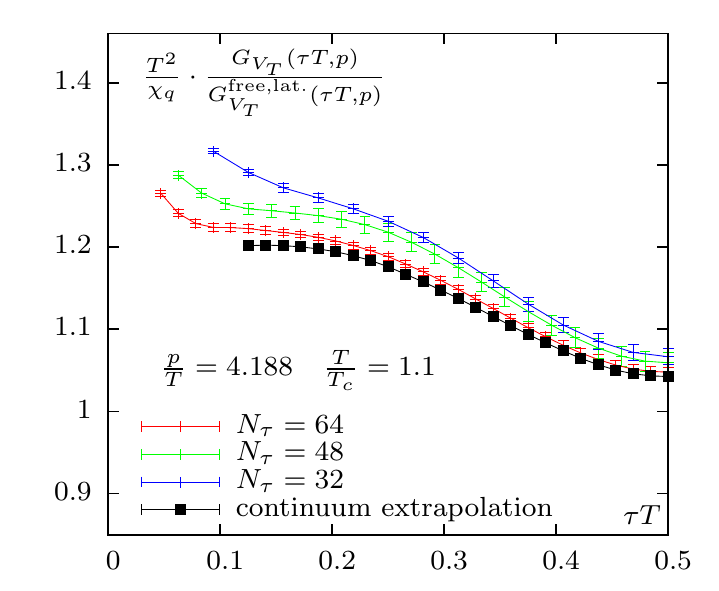}
\includegraphics{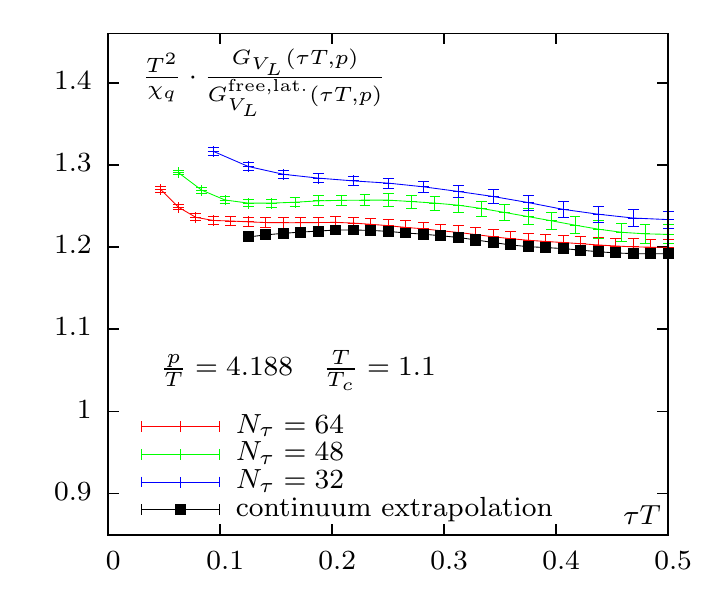}
\caption{Transverse~(left) and longitudinal~(right) component of the vector
  correlation function at $1.1~T_c$ and momentum $p/T=4.188$.}
\label{plot:momenta_lat}
\end{center}
\end{figure}

\begin{equation}
\frac{\sigma}{T} = \frac{C_\text{em}}{6} \lim_{\omega\rightarrow 0} \frac{\rho_{ii}(\omega)}{\omega T}
\quad \rightarrow \quad \sigma(T) / C_\text{em} = 2 \chi_q c_\text{BW} / (3 \Gamma)
\label{elcon}
\end{equation}
where $C_\text{em}$ is given by the
elementary charges $Q$ of the quark flavor $f$ 
as $C_\text{em}=\sum_f Q_f^2 $, 
and the thermal production rate of dilepton pairs as

\begin{equation}
\frac{dW}{d\omega d^3 p} = \frac{5 \alpha^2}{54 \pi^3 } 
\frac{1}{\omega^2 ( e^{\omega / T} - 1 )} 
\label{dileptonrate}
\rho_{ii}(\omega,p,T).  
\end{equation}

\section{Results for correlators and spectral function fits}

We have calculated vector correlation functions using non-perturbatively
clover-improved Wilson fermion on a set of quenched gauge field configurations
stated in Tab.~\ref{simulationParameters}. 
For each temperature three lattice spacings were used in order to perform the
continuum limit.
While the results of \cite{Ding:2010ga} were obtained on a fixed spatial lattice of
$N_\sigma=128$, recent results were calculated at fixed aspect ratio of
$N_\sigma/N_\tau=3.00$ for 1.1~$T_c$ and $N_\sigma/N_\tau=3.43$ at
1.2~$T_c$ and therefore at a fixed physical volume. This set-up allows for a
continuum extrapolation also at finite
momenta for the two lowest temperatures.
In Fig.~\ref{plot:V1234c} the results of the vector correlation
function $G_{ii}(\tau T)$ for 1.1~$T_c$~(left) and 1.2~$T_c$
(right) are shown together with the corresponding continuum extrapolations. 
All results are normalized by the free non-interacting
correlation function, $G_V^{\text{free,lat}}(\tau T)$, and with the quark number
susceptibility, $\chi_q$, to remove any renormalization constants.
While the results on finite lattices show strong cut-off effects at all
distances and a wrong rising behavior towards small distances, the continuum
extrapolations are well behaved down to distances around $0.1\tau T$. 
The band shows the result and systematic error of the fit using the Ansatz
(\ref{spfAnsatz}).
The rise
of these ratios towards the mid-point can be identified as a first indication
for a transport contribution. 

The resulting spectral functions using the Ansatz (\ref{spfAnsatz})
are shown in Fig.~\ref{plot:spf}~(left) and the dilepton rates obtained from 
(\ref{dileptonrate}) in Fig.~\ref{plot:spf}~(right).
Both results are compared to a dilepton spectrum calculated within the hard
thermal loop approximation \cite{Braaten:1989mz}.
In the limit of vanishing $\omega$ one can read of the
electrical conductivity (\ref{elcon}) from Fig.~\ref{plot:spf}~(left). The results
for $\sigma/T$ plotted in
Fig.~\ref{plot:cond} show no temperature dependence in the analyzed T-range
within the systematic errors. 
For a comparison with results from calculations with dynamical quarks on finite
lattices see \cite{Amato:2013naa,Brandt:2012jc}.

\section{Vector correlation function at non-vanishing momenta}

\begin{figure}
\begin{center}
\includegraphics{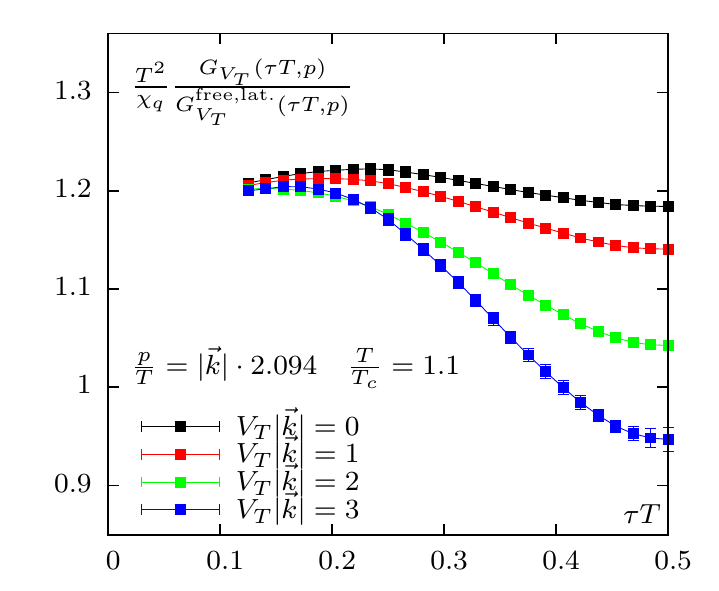}
\includegraphics{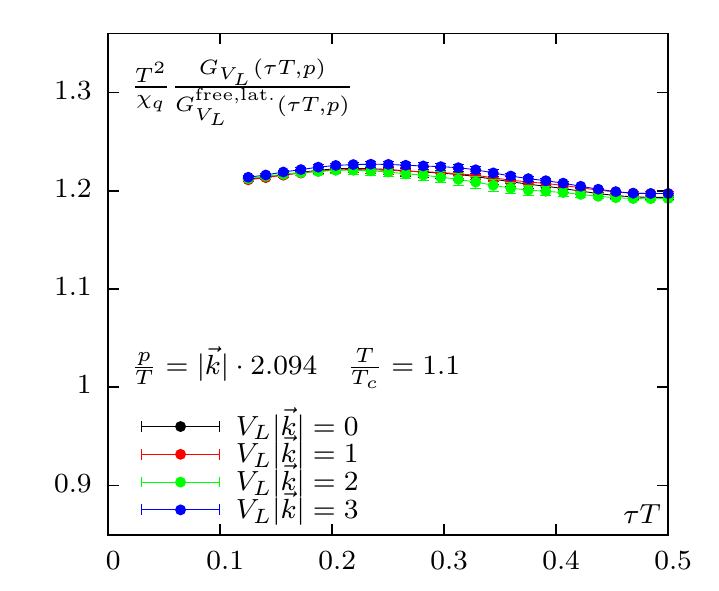}
\caption{Continuum extrapolated vector correlation function in transversal
 ~(left) and longitudinal~(right) polarization $1.1~T_c$ and four momenta.}
\label{plot:momenta_cont}
\end{center}
\end{figure}

For the two lowest temperatures, 1.1~$T_c$ and 1.2~$T_c$, the vector correlation
functions for the three lattice spacings were calculated at a fixed aspect
ratio of $N_\sigma/N_\tau$. This allows to perform the continuum extrapolation
also for non-vanishing momenta, $\vec p$. In Fig.~\ref{plot:momenta_lat} and
Fig.~\ref{plot:momenta_cont} the results for the
vector correlation functions in the transversal~(left) and longitudinal~(right) polarization
channel at a temperature of 1.1~$T_c$ are shown. 
While for the longitudinal
channel no momentum dependence is observed within the errors, the transversal
channel shows a pronounced $\vec p$-dependence which is weak at small
separations but becomes stronger at larger distances towards the mid-point.\\
This behavior can qualitatively be interpreted in the context of the
small-frequency behavior of the two channels in the
leading-log Boltzmann approximation \cite{Hong:2010at}. In the limit
of $\omega\rightarrow 0$ the transverse spectral function
$\rho_T(\omega,\vec p)/\omega$ is non-vanishing for all momenta $\vec p$, while it is
zero in the longitudinal polarization, $\rho_T(\omega,\vec p)/\omega$, for all non-vanishing momenta. This leads to a different $\vec
p$-dependence of the low-frequency region which results in a different
dependence on momenta of the correlation functions at large distances and a
more pronounced $\vec p$-dependence in the transversal channel.
Due to asymptotic freedom one expects that both channels approach each other at
small distances.
Based on these observations, the different behavior in the transversal and
longitudinal polarization can qualitatively be understood,
but for a more detailed understanding the spectral properties need to be
determined in the future.
For a comparison of our continuum extrapolated vector correlation function at
non-vanishing momenta with a perturbative NLO calculation see \cite{Laine:2013vma}.

\section{Conclusion}
Continuum results for the vector current correlation functions in
the temperature ranges $1.1\leq T/T_c \leq 1.4~T_c$ were presented. Using a
rather simple, phenomenologically inspired Ansatz for the vector spectral
function and a careful analysis of the systematic uncertainties, this sheds
light into the temperature dependence of the electrical conductivity and
dilepton rates in an energy regime currently accessible in heavy-ion
experiments. The continuum extrapolated vector correlation function at
non-vanishing momenta show a qualitatively different behavior in the
transversal and longitudinal polarization. The results allow for study of the
spectral properties and determination of the photon rates in the future.

\section*{Acknowledgment}

The results have been
achieved by using the PRACE Research
Infrastructure resource JUGENE based at the
J\"ulich Supercomputing Centre in Germany
and the Bielefeld GPU-cluster resources.
This work has been partly supported by the
IRTG/GRK 881 "Quantum Fields and Strongly Interacting Matter".


\begin{thebibliography}{99}

%%\cite{Ding:2012sp}
%\bibitem{Ding:2012sp} 
%  H.~T.~Ding et al.,
%  %``Charmonium properties in hot quenched lattice QCD,''
%  Phys.\ Rev.\ D {\bf 86}, 014509 (2012).
%  %%[arXiv:1204.4945 [hep-lat]].
%  %%CITATION = ARXIV:1204.4945;%%
%  %31 citations counted in INSPIRE as of 25 Nov 2013

\bibitem{Rapp:2009yu}
R.~Rapp, J.~Wambach, and H.~van Hees, {\it {The Chiral Restoration Transition
  of QCD and Low Mass Dileptons}},  in {\em Landolt-B{\"o}rnstein}, vol.~I-23,
  4-1.
\newblock Springer-Verlag, 2010.
\newblock \href{http://arxiv.org/abs/0901.3289}{{\tt arXiv:0901.3289}}.

%\bibitem{pos-olaf}
%O.~Kaczmarek, 
%%{\it {Recent Developments in Lattice Studies for Quarkonia}},
%{\em Nucl.~Phys.~A~910} (2012) 98.
%%[{{\tt arXiv:1208.4075}}].

%\bibitem{pos-hengtong}
%H.-T. Ding, 
%%{\it In-medium hadron properties from lattice qcd},  
%{\em EPJ Web  of Conferences} {\bf 36} (2012) 00008.
%%  [\href{http://arxiv.org/abs/1207.5187}{{\tt arXiv:1207.5187}}].

\bibitem{Ding:2010ga}
H.-T. Ding, A.~Francis, O.~Kaczmarek, F.~Karsch, E.~Laermann, et~al., 
%{\it
%  {Thermal dilepton rate and electrical conductivity: An analysis of vector
%  current correlation functions in quenched lattice QCD}}, 
{\em Phys.Rev.}  {\bf D83} (2011) 034504.
% [\href{http://arxiv.org/abs/1012.4963}{{\tt
%  arXiv:1012.4963}}].

\bibitem{Francis:2011bt}
A.~Francis and O.~Kaczmarek, 
%{\it {On the temperature dependence of the
%  electrical conductivity in hot quenched lattice QCD}},  
 {\em
  Prog.Part.Nucl.Phys.} {\bf 67} (2012) 212.
%  [\href{http://arxiv.org/abs/1112.4802}{{\tt arXiv:1112.4802}}].

%\cite{Ding:2013qw}
\bibitem{Ding:2013qw} 
  O.~Kaczmarek et al.,
  %``Thermal dilepton rates from quenched lattice QCD,''
  PoS ConfinementX (2012) 185.
%%  [arXiv:1301.7436 [hep-lat]].
  %%CITATION = ARXIV:1301.7436;%%
  %8 citations counted in INSPIRE as of 29 Nov 2013

\bibitem{Allton:2005gk}
C.~Allton, M.~Doring, S.~Ejiri, S.~Hands, O.~Kaczmarek, et~al.,
% {\it
%  {Thermodynamics of two flavor QCD to sixth order in quark chemical
%  potential}},  
{\em Phys.Rev.} {\bf D71} (2005) 054508.
%  [\href{http://arxiv.org/abs/hep-lat/0501030}{{\tt hep-lat/0501030}}].

\bibitem{Gavai:2001ie}
R.~V. Gavai, S.~Gupta, and P.~Majumdar, 
%{\it {Susceptibilities and screening
%  masses in two flavor QCD}},  
{\em Phys.Rev.} {\bf D65} (2002) 054506.
%  [\href{http://arxiv.org/abs/hep-lat/0110032}{{\tt hep-lat/0110032}}].

\bibitem{Aarts:2002cc}
G.~Aarts and J.~M. Martinez~Resco, 
%{\it {Transport coefficients, spectral
%  functions and the lattice}},  
{\em JHEP} {\bf 0204} (2002) 053.
%  [\href{http://arxiv.org/abs/hep-ph/0203177}{{\tt hep-ph/0203177}}].

\bibitem{Moore:2006qn}
G.~D. Moore and J.-M. Robert, 
%{\it {Dileptons, spectral weights, and
%  conductivity in the quark-gluon plasma}},
\href{http://arxiv.org/abs/hep-ph/0607172}{{\tt hep-ph/0607172}}.

\bibitem{Petreczky:2005nh}
P.~Petreczky and D.~Teaney, 
%{\it {Heavy quark diffusion from the lattice}},
  {\em Phys.Rev.} {\bf D73} (2006) 014508.
%  [\href{http://arxiv.org/abs/hep-ph/0507318}{{\tt hep-ph/0507318}}].

\bibitem{Hong:2010at}
J.~Hong and D.~Teaney, 
%{\it {Spectral densities for hot QCD plasmas in a
%  leading log approximation}},  
{\em Phys.Rev.} {\bf C82} (2010) 044908.
%  [\href{http://arxiv.org/abs/1003.0699}{{\tt arXiv:1003.0699}}].


\bibitem{Burnier:2012ts}
Y.~Burnier and M.~Laine, 
%{\it {Towards flavour diffusion coefficient and
%  electrical conductivity without ultraviolet contamination}},  
{\em
  Eur.Phys.J.} {\bf C72} (2012) 1902.
%  [\href{http://arxiv.org/abs/1201.1994}{{\tt arXiv:1201.1994}}].

%\cite{Rapp:2013nxa}
\bibitem{Rapp:2013nxa} 
  R.~Rapp,
  %``Dilepton Spectroscopy of QCD Matter at Collider Energies,''
  arXiv:1304.2309 [hep-ph].
  %%CITATION = ARXIV:1304.2309;%%
  %4 citations counted in INSPIRE as of 29 Nov 2013

\bibitem{Braaten:1989mz}
E.~Braaten and R.~D. Pisarski, 
%{\it {Soft Amplitudes in Hot Gauge Theories: A
%  General Analysis}},  
{\em Nucl.Phys.} {\bf B337} (1990) 569.

%\cite{Amato:2013naa}
\bibitem{Amato:2013naa} 
  A.~Amato et al.,
  %``Electrical conductivity of the quark-gluon plasma across the deconfinement transition,''
  Phys.\ Rev.\ Lett.\  {\bf 111}, 172001 (2013).
%  [arXiv:1307.6763 [hep-lat]].
  %%CITATION = ARXIV:1307.6763;%%
  %12 citations counted in INSPIRE as of 27 Nov 2013

%\cite{Brandt:2012jc}
\bibitem{Brandt:2012jc} 
  B.~B.~Brandt, A.~Francis, H.~B.~Meyer and H.~Wittig,
  %``Thermal Correlators in the \rho\ channel of two-flavor QCD,''
  JHEP {\bf 1303}, 100 (2013).
  %[arXiv:1212.4200 [hep-lat]].
  %%CITATION = ARXIV:1212.4200;%%
  %12 citations counted in INSPIRE as of 29 Nov 2013

%\cite{Laine:2013vma}
\bibitem{Laine:2013vma} 
  M.~Laine,
  %``NLO thermal dilepton rate at non-zero momentum,''
  JHEP {\bf 1311}, 120 (2013).
%  [arXiv:1310.0164 [hep-ph]].
  %%CITATION = ARXIV:1310.0164;%%
  %1 citations counted in INSPIRE as of 27 Nov 2013

%\bibitem{Aarts:2007wj}
%G.~Aarts, C.~Allton, J.~Foley, S.~Hands, and S.~Kim, 
%%{\it {Spectral functions
%%  at small energies and the electrical conductivity in hot, quenched lattice
%%  QCD}},  
%{\em Phys.Rev.Lett.} {\bf 99} (2007) 022002.
%%  [\href{http://arxiv.org/abs/hep-lat/0703008}{{\tt hep-lat/0703008}}].

%%%%%%%%%%%%%%%%% New


\end{thebibliography}
\end{document}